\documentclass[11pt,a4paper]{article}
\usepackage[hyperref]{naaclhlt2019}
\usepackage{times}
\usepackage{graphicx}
\usepackage{latexsym}
\usepackage{bbm}
\usepackage{booktabs}
\usepackage[final]{pdfpages}

\usepackage{url}

\aclfinalcopy 

\title{Simple dynamic word embeddings for mapping perceptions in the public sphere}

\author{Nabeel Gillani \\
  MIT \\
  {\tt ngillani@mit.edu} \\\And
  Roger Levy \\
  MIT \\
  {\tt rplevy@mit.edu} \\}

\date{}

\begin{document}
\maketitle
\noindent\fbox{%
    \parbox{\linewidth}{%
        This is a modified version of a paper that originally appeared in the 2019 NAACL workshop on NLP+CSS.  Since its initial publication, we discovered implementation errors that invalidate the results presented in the original version of the paper.  These errors are corrected in this version.  A corrigendum at the end of this paper summarizes the original errors and how we corrected them.
    }%
}
\begin{abstract}
    Word embeddings trained on large-scale historical corpora can illuminate human biases and stereotypes that perpetuate social inequalities.  These embeddings are often trained in separate vector space models defined according to different attributes of interest.  In this paper, we develop a unified dynamic embedding model that learns attribute-specific word embeddings.  We apply our model to investigate i) 20th century gender and ethnic occupation biases embedded in the Corpus of Historical American English (COHA), and ii) biases against refugees embedded in a novel corpus of talk radio transcripts containing 119 million words produced over one month across 83 stations and 64 cities.  Our results shed preliminary light on scenarios when dynamic embedding models may be more suitable for representing linguistic biases than individual vector space models, and vice-versa.
\end{abstract}
\section{Introduction}

Language has long been described as both a cause and reflection of our psycho-social contexts~\cite{lewisLanguageShapesCulture}.  Recent work using word embeddings---low-dimensional vector representations of words trained on large datasets to capture key semantic information---has demonstrated that language encodes several gender, racial, and other biases that correlate with both implicit biases~\cite{caliskan2017semantics} and historical trends~\cite{gargEmbeddingsStereotypes}.  

These studies have validated the use of word embeddings to measure a range of psychological and social contexts, yet in most cases, they do not leverage the full power of available datasets.  For example, the historical biases presented in~\cite{gargEmbeddingsStereotypes} are computed  using decade-specific word embeddings produced by training different Word2Vec~\cite{mikolovW2V} models on a large corpus of historical text from that decade.  The authors then use a Procrustes alignment to project embeddings from different models into the same vector space so they can be compared across decades~\cite{hamiltonHistWords}.  While this approach is reasonable when there are large-scale datasets available for a given attribute of interest (e.g. decade), it requires an additional optimization step and disregards valuable training data that could be pooled and leveraged across attribute values to help with both training and regularization---especially when data is sparse.

In this paper, we present a unified dynamic word embedding model that jointly trains linguistic information alongside any categorical variable of interest describing its context (-e.g. year, geography, income bracket, etc.).  We apply this model to two datasets: i) the Corpus of Historical American English (COHA~\cite{daviesCOHA}) to analyze gender and ethnic occupation biases, and ii) a novel data corpus of 119 million words spoken on talk radio~\cite{beefermanTalkRadio} during a one-month period in late 2018 across 64 US cities to explore perceptions about refugees.  Our results shed preliminary light on scenarios when dynamic embedding models may be more suitable for representing linguistic biases than individual vector space models, and vice-versa.

\section{Model}
We describe our model and implementation below.
\subsection{Overview}
Our dynamic embedding for word $w$ is defined as

\begin{equation}
    E(w, A) = \gamma_w + \Sigma_{a\in A}~\beta_w^{a} \label{eq:dynamic-embedding}
\end{equation}

\noindent
where $\gamma_w$ is an attribute-invariant embedding of $w$ computed across the entire corpus, $\beta_w^{a}$ is the offset for $w$ with respect to attribute $a$ across the set of attributes $A$ we are interested in computing the word embedding with respect to.  For example, if we wish to compute the embedding for the word ``refugee'' as it was used on the 25th day of a particular 30-day corpus of talk radio transcripts, we would set $w=refugee$ and $A=\{25\}$.  This approach, as formalized in Equation~\ref{eq:dynamic-embedding} above, is identical to one introduced by~\cite{bammanGeoEmbedding}, though finer details of our model and training differ slightly, as described below.

To learn $\gamma_w$ and $\beta_w^{a}$, we train a neural network.  Our model is a simple extension to the distributed memory (DM) model for learning paragraph vectors originally introduced in~\cite{leParagraphVectors}.  The DM model uses a continuous bag-of-words architecture to jointly train a paragraph ID with a sequence of words sampled from that paragraph to predict a particular word given the words that surround it.  The output of this model includes a semantic vector representation of a) each paragraph, and b) each word in the vocabulary.

\begin{figure}
\includegraphics[width=1.0\linewidth]{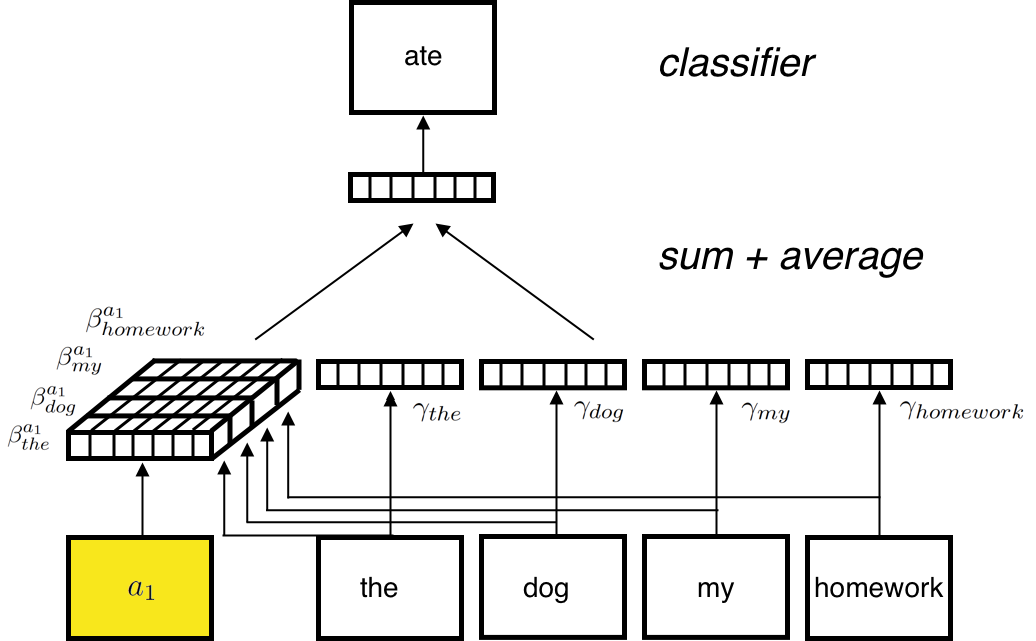}
\caption{Our dynamic embedding model learns an attribute invariant embedding for each training word $w$ (i.e., $\gamma_w$), along with an attribute-specific offset for attribute $A = \{a_1\}$ (i.e., $\beta_w^{a_1}$).  The $\gamma_w$ and $\beta_w^{a_1}$ terms are summed to compute $E(w,A)$ for each context word and averaged across words before classification.  Figure inspired by~\cite{leParagraphVectors}.}
\label{fig:model}
\end{figure}

Our model extends the DM model by adding an additional dimension to the paragraph vector to learn specific \textit{paragraph-by-word}---or, in our context, \textit{attribute-by-word}---embeddings (i.e., $\beta_w^{a}$).  The penultimate layer (before word prediction) is computed as an average of the dynamic embeddings for each context word, i.e., $X = \frac{1}{N}\Sigma_{i=1}^N E(w_i, S, A) $, where $N$ is the size of our context window.  This average embedding is then multiplied by the output layer parameters and fed through the final layer for word prediction.  Figure \ref{fig:model} depicts our model architecture.  

\subsection{Implementation}
We build on an existing PyTorch implementation of paragraph vectors\footnote{Available at: \url{https://github.com/inejc/paragraph-vectors}.} to implement our model, setting the dimensionality of $\gamma_w$ and $\beta_w^{a}$ to be 100.  We use the Adam optimization algorithm with a batch size of 128, word context window size of 8 (sampling four words to the left and right of a target prediction word), learning rate of 0.001, and L2 penalty to regularize all model parameters (where $\lambda$=1e-5).  We only train embeddings for words that occur at least 10 times in the corpus.  For training, we use the negative sampling loss function, which~\cite{mikolovW2V} show is more efficient than the hierarchical softmax and yields competitive results\footnote{We include three noise words when computing the loss.}.

\section{Case study 1: gender and ethnic occupation biases in COHA}
We first train our model on the Corpus of Historical American English~\cite{daviesCOHA} for 2 epochs (each epoch takes approximately 40 hours to train) in order to compare its outputs to those produced via the individual decade-by-decade word embedding models used in~\cite{gargEmbeddingsStereotypes}.  We use the same metric and word lists as the authors to compute bias scores, substituting in the cosine distances between vectors for the norm of their difference (both approaches yield nearly identical results).   In particular, we compute linguistic bias scores for two of their analyses: i) the extent to which female versus male words are semantically similar to occupation-related words, and ii) the extent to which Asian vs.\ White last names are semantically similar to the same, from 1910 through 1990.  We then qualitatively analyze relationships between changes in these scores and the actual changes in female and Asian workforce participation rates (relative to men and Whites, respectively) over the same time period.  

\begin{figure*}
\includegraphics[width=\linewidth]{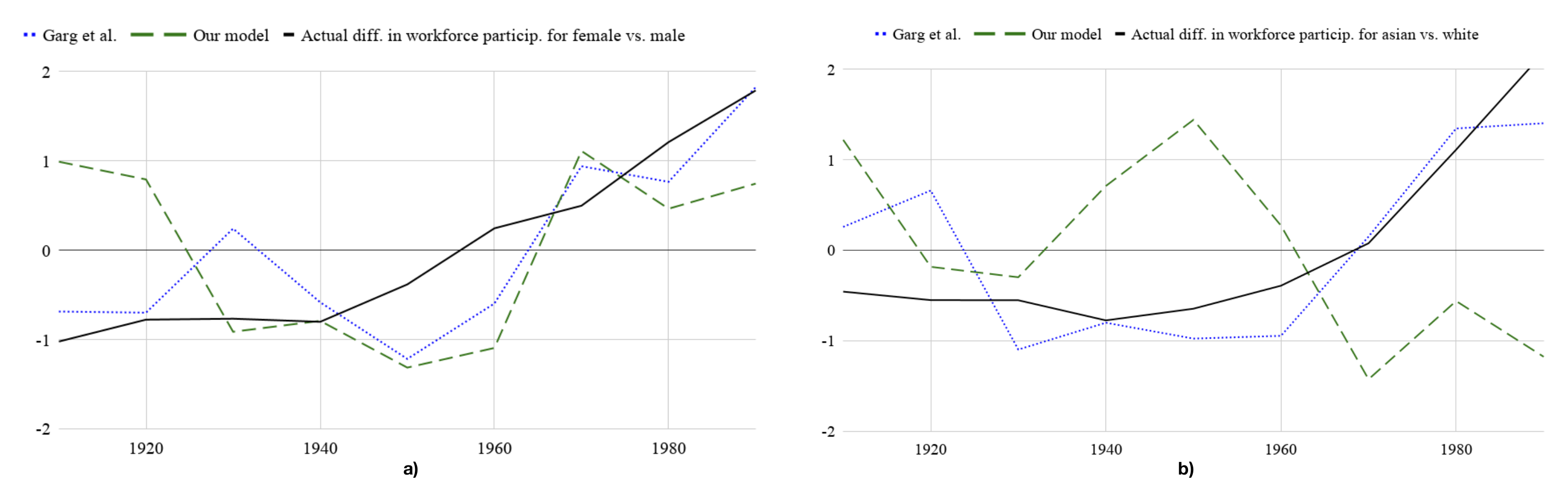}
\caption{Scores produced by~\cite{gargEmbeddingsStereotypes} and our model (blue dotted and green dashed lines, respectively) compared to actual workforce participation rates (solid lines) for gender (top) and Asian/White (bottom) linguistic biases.  To compare all values on a single y-axis, we standardize both sets of bias scores and workforce participation rates by subtracting the mean and dividing by the standard deviation across decades.}
\label{fig:validation}
\end{figure*} 

Figure \ref{fig:validation} depicts these results.  The results from the dynamic model depicted in (a) qualitatively appear to match the ground truth workforce participation rates, except in the earlier decades of the 20th century.  The results depicted in (b), however, do not appear to resemble those in~\cite{gargEmbeddingsStereotypes} or ground truth.  One hypothesis for these trends, especially those depicted in (b), is that the attribute value-specific offset word vectors might not be encoding the right, or enough, information about those words when compared to the word's attribute-invariant embedding.  This seems plausible especially since the norms of the attribute-invariant word vectors are often one to two orders of magnitude larger than the norms of their corresponding attribute value-specific offsets.  Given research that has identified that words occurring more frequently in consistent semantic contexts tend to have larger norms~\cite{schakelWordSig}, it is possible that some of the Asian last names used as input words into the computations for (b) simply do not occur frequently enough within the individual decades for their decade-specific representations to retain explanatory power when compared to their corpus-wide, attribute-invariant representations.  An analogous reason might explain why the bias scores for the earlier decades in the 20th century are so high: it is possible that the corpus-wide embeddings for certain words used in those calculations are simply ``overpowering'' their decade-specific offsets. 

\begin{figure*}
\includegraphics[width=\linewidth]{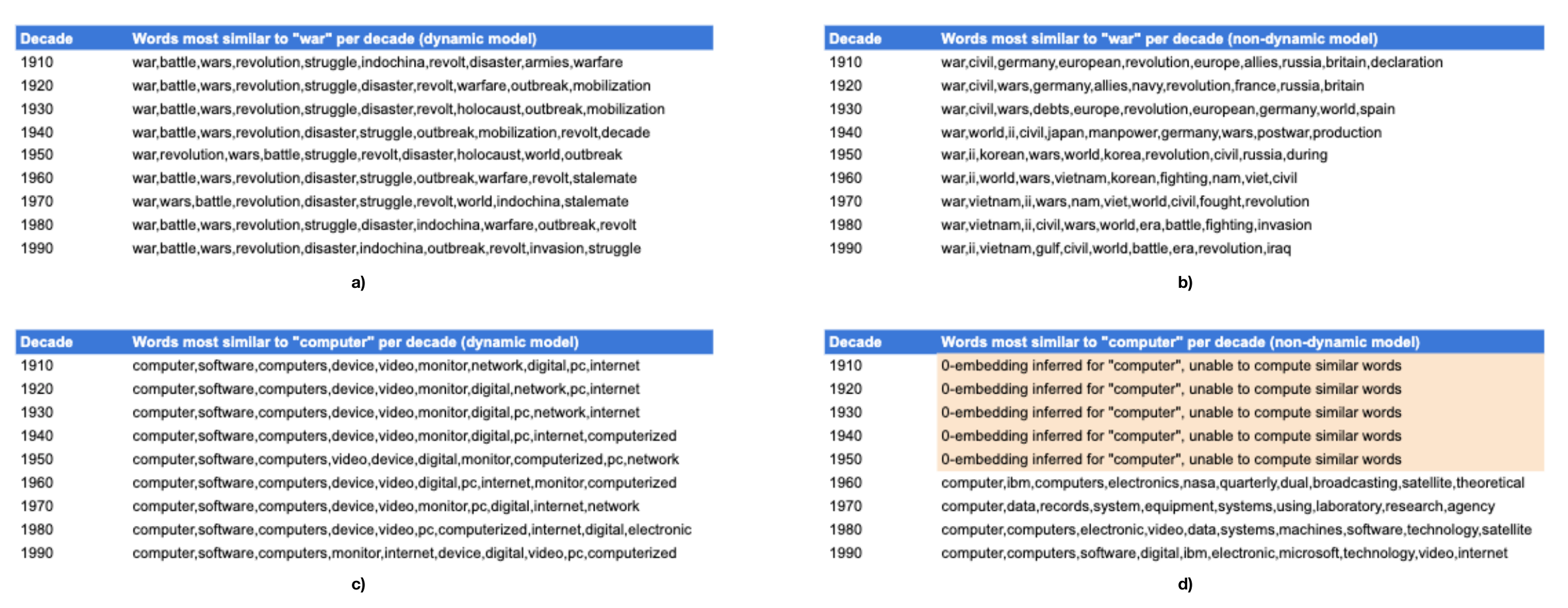}
\caption{Words most similar to ``war'' and ``computer'' when using our dynamic model (a) and c)) and the non-dynamic model from~\cite{gargEmbeddingsStereotypes} (b) and d)), respectively.}
\label{fig:decade_by_decade}
\end{figure*}

Running some additional tests starts to confirm some of these hypotheses.  For example, ``war'' was a prominent feature of the 20th century, and so, we might expect the most semantically-similar words per decade to the word ``war'' to relate to the wars that were particularly salient during those (or adjacent) decades.  Figure~\ref{fig:decade_by_decade}(a) shows (in order of cosine similarity) the words per decade that are most similar to that decade-specific embedding for ``war''.  We can see that in some cases, our intuition holds: e.g., the 1950s contain words like ``holocaust'' and ``world'' (presumably for``world war''), both of which seem to be reasonable for that time period.  However, ``holocaust'' also appears in the list for 1930---even though the term was first mentioned in the New York Times in the context of the second World War in the early 1940s\footnote{\url{https://en.wikipedia.org/wiki/Names_of_the_Holocaust}.}.  Possible explanations for these observations include: i) there is a strong corpus-wide association between ``war'' and ``holocaust'' that makes them similar across decades (even those preceding the Holocaust), and ii) the 1930s-specific offset for ``holocaust'' was perhaps simply not updated enough during training (due to a low number of occurrences) to move away from its initialized (random values).  If those initialized values were small enough (e.g. had a small norm), it's quite possible that ``holocaust'' could still show up as a semantically similar term to ``war'' even before it became a more commonly-used term.  These appear to be plausible explanations when we look at the decade-specific associations with ``war'' computed using the non-dynamic model of~\cite{gargEmbeddingsStereotypes} and shown in Figure~\ref{fig:decade_by_decade}(b): the associations appear to depict historical realities much more accurately than those inferred by our dynamic model.


We see a similar pattern when looking at words that are most similar to ``computer'' per decade. Figure~\ref{fig:decade_by_decade}(c) shows words that are most similar according to our dynamic model, and (d) shows the most similar words according to the non-dynamic model.  According to our dynamic model, ``internet'' is a nearest neighbor as early as 1910, which is clearly historically inaccurate.  On the other hand, the non-dynamic model does not learn an embedding for computer until the 1960s (due to limited use of the word in the first half of the 20th century), after which it produces similar words that appear to accurately reflect decade-specific trends. 

One opportunity for future work involves trying different (smaller) values of $\lambda$ in our l2 regularization term, as doing so might encourage attribute value-specific embeddings for a given word to retain more ``influence'' in its overall representation.  In any case, our results reveal some of the possible shortcomings of using our dynamic embedding model, particularly when there is enough data to simply learn and align individual attribute-specific word embedding models.  

\section{Case study 2: bias against refugees expressed on talk radio}
The earlier section revealed some of the shortcomings of our model when applied to a large historical corpus.  To explore its performance in a smaller-scale corpus captured over a much shorter time horizon, we apply our model to the transcriptions of talk radio shows in order to identify biases against refugees.  Talk radio is a significant source of news for a large fraction of Americans: In 2017, over 90\% of Americans over the age of 12 listened to some type of broadcast radio during the course of a given week, with news/talk radio serving as one of the most popular types~\cite{pewRadioStats}.  With listener call-ins and live dialog, talk radio provides an interesting source of information, commentary, and discussion that distinguishes it from discourse found in both print and social media.  Given the proliferation of refugees and displaced peoples in recent years (totalling nearly 66 million individuals in 2016~\cite{unhcrStats})---coupled with the rise of talk radio as a particularly popular media channel for conservative political discourse~\cite{mortDissentAirwaves}---analyzing bias towards refugees across talk radio stations may provide a unique window into how Americans perceive this important issue.

\subsection{Dataset and analyses}
Our data is sourced from talk radio audio data collected and automatically transcribed by the media analytics nonprofit Cortico\footnote{\url{http://cortico.ai}.}.  The data is further processed to identify different speaker turns into ``snippets''; infer the gender of the speaker; and compute other useful metrics (more details on the radio data pipeline can be found in~\cite{beefermanTalkRadio}).  

We train our dynamic embedding model on a talk radio datasets sourced from 83 stations located in 64 cities across the US.  The dataset includes over 4.8 million snippets comprised of 119 million total words produced by 433 shows between August 15, and September 15, 2018\footnote{As a rough proxy for removing syndicated content, we include only those snippets produced by a talk radio shows that air on one station.}.

Bias against refugees is defined similarly to how the authors of~\cite{gargEmbeddingsStereotypes} define bias against Asians during the 20th century, measuring to what extent radio shows associate ``outsider'' adjectives like ``aggressive'', ``frightening'', ``illegal'', etc. with refugee and immigrant-related terms in comparison to all other adjectives.  To compute refugee bias scores with respect to the attribute set $A$, we use a modified version of the relative norm distance metric from~\cite{gargEmbeddingsStereotypes}:

\[
    bias_A = \frac{\Sigma_{r\in R}~cos(E(r, A), \overline{a}) - cos(E(r, A), \overline{o})}{|R|}
\]

Where $E(r, A)$ is the dynamic embedding for a given refugee-related word $r$ (e.g. ``refugee'', ''immigrant'', ''asylum'', etc); $\overline{a}$ is the average dynamic embedding computed for each $w$ in the set of all adjectives with respect to $A$; $\overline{o}$ is analogously defined for outsider adjectives; and $cos(\cdot)$ is cosine distance.  A positive value for $bias_A$ indicates discourse about refugees that is biased against them as ``outsiders''.  We normalize by the total number of refugee-related words, $R$, to provide a relative indication of the amount of linguistic bias present in the data (since $bias_A$ is bounded by $\pm|R|$).
 
\begin{figure*}[ht]
\includegraphics[width=\linewidth]{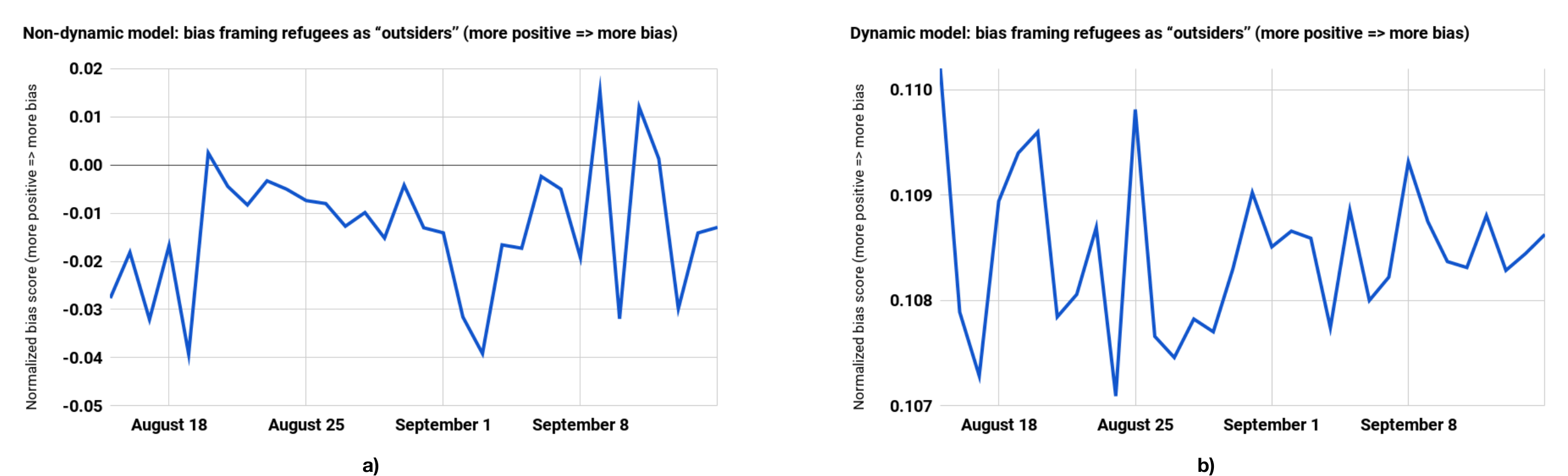}
\caption{Bias towards refugees as outsiders across talk radio shows from mid-August to mid-September 2018: (a) depicts bias scores computed using a ``non-dynamic model'', i.e., training multiple Word2Vec models (one per day of data) and then projecting these models into the same vector space using orthogonal Procrustes alignment, and (b) depicts bias scores computed using our dynamic model.}
\label{fig:refugee_bias_time}
\end{figure*}

We use our model to analyze how bias on talk radio against refugees varies by day between August 15 and September 15, 2018.  The median number of words for each day in the talk radio corpus is 4 million---over 5x fewer than a median of 22 million words per decade used to train each decade-specific model in~\cite{gargEmbeddingsStereotypes}.  As a comparison, we also compute bias scores by training one Word2Vec model per day and projecting all day-by-day models into the same vector space using orthogonal Procrustes alignment\footnote{We use the Gensim implementations of Word2Vec and orthogonal Procrustes alignment, aligning hyperparameters as closely as possible to our dynamic model.} similar to~\cite{hamiltonHistWords}.  The results from these analyses are depicted in Figure~\ref{fig:refugee_bias_time}.  Unlike the earlier section where we used actual workforce participation rates as a ``ground truth'' to compare linguistic biases against, it is unclear what ground truth is in this case in order to be able to evaluate model performance.  Still, one key difference in the results is that most of the daily bias scores computed using the non-dynamic model are negative---suggesting, in general, that talk radio participants do not discuss refugees as ``outsiders''---whereas all of the scores computing using the dynamic model are positive.  One reason to believe the latter is a more accurate depiction of reality---that discourse about refugees on talk radio would tend to cast them as ``outsiders''---is that talk radio has been identified as a popular media source for political conservatives in the US~\cite{mortDissentAirwaves}.  In turn, political conservatives---and particularly their more extreme right-wing factions---have historically been unwelcoming towards refugees~\cite{bencekRefugeeViolence,klausRefugeePolicy}.   

A possible analogue to using longer-term historical changes in workforce participation participation rates as a proxy for ``ground truth'' is to use shorter-term ``historical'' changes in refugee-related news events.  Upon qualitative inspection, it appears that the results from our dynamic embedding model might be tracking the news cycle.  For example, the lowest bias scores in the chart---August 17 and 24---could perhaps correspond to stories about how ``U.S. Will Not Spend \$230 Million Allocated to Repair Devastated Syrian Cities'' and ``For Rohingya, Years of Torture at the Hands of a Neighbor'', respectively (as reported by the New York Times).  Intuitively, public discourse that is relatively less-biased against (and perhaps even more empathetic towards) refugees in response to these stories appears to make sense. Conversely, the highest bias scores on August 15 and 25 could be related to other New York Times stories titled ``ISIS Member Arrested in Sacramento, U.S. Says''\footnote{in which case, the arrested individual had actually applied for refugee status in the US}, and ``Year After Rohingya Massacres, Top Generals Unrepentant and Unpunished'', respectively.  In the case of the latter story, it seems much of the radio discussion describes ``angry protests'' by refugees affected by the Rohingya crisis---which may be contributing to elevated negative/``outsider'' bias scores.  Of course, there are several seemingly prominent refugee-related events that do not correspond to extreme daily bias scores, e.g. the September 14 announcement that the ``U.S. Is Ending Final Source of Aid for Palestinian Civilians''. 

While the non-dynamic model's mostly-negative absolute scores might suggest that it fails to capture generally unwelcoming discourse about refugees, its extreme peaks and valleys, too, correspond to news stories in ways we might expect.  For example, the two days with the lowest bias scores---August 19 and September 3---coincide with stories titled "‘All of Africa Is Here’: Where Europe’s Southern Border Is Just a Fence" and "Mediterranean Death Rate Is Highest Since 2015 Migration Crisis", respectively.  Both of these articles describe the treacherous journeys many refugees embark on as they flee their countries.  One of the days with the highest bias scores (September 9) coincides with a story titled "Sweden’s Centrists Prevail Even as Far Right Has Its Best Showing Ever" (which also references the right-wing party's anti-immigrant positions); the other (September 11) does not appear to correspond to a specific refugee-related story.  

Together, these results suggest that our dynamic model may produce embeddings that more accurately capture linguistic biases towards refugees, but also that using the news cycle as ``ground truth'' is not sufficient for evaluation purposes.  Future research should include i) more thoughtful selections of ``ground truth'' for model comparisons and validation, ii) comparisons to other dynamic word embedding models that treat time as a continuously-valued attribute, e.g.~\cite{bamlerDynamicWE,rudolphLanguageEvolution,yaoSemanticDiscovery}, and iii) an exploration of if/how these findings hold for topics other than refugees, across different time slices and discourse corpora.\footnote{In the original version of this paper, we also included a case study where we analyzed how bias against refugees differs by geography.  This was a largely speculative analysis that we included after validating our model using COHA and the time-series of radio data discussed in this section.  Since our model's validity is much weaker in this corrected version of the paper, we have removed this speculative geographic analysis.  Additional details on the original analyses can be found in the corrigendum.}




\section{Conclusion}
In this paper, we present a dynamic word embedding model mirroring the earlier work of~\cite{bammanGeoEmbedding} to learn attribute-specific embeddings.  We observe several limitations of our model in comparison to the non-dynamic models used in~\cite{gargEmbeddingsStereotypes} to measure gender and ethnic occupation biases embedded in COHA.  However, preliminary results suggest our model might more accurately capture linguistic biases expressed against refugees on talk radio, where data per attribute value is much sparser than in COHA.  Our results are highly preliminary and require further investigation to determine under which conditions dynamic embedding models like ours might be more suitable to use than individual attribute-specific models.  Some directions for investigation include: i) training our model with different hyperparameter configurations---especially changes in $\lambda$ for l2 regularization to assess its impact on the influence of words' attribute-specific offsets on their overall embeddings; ii) exploring applications to different datasets and topics; iii) more thoughtful selections for ``ground truth'' to evaluate results, and iv) comparing the results of our model to other temporal embedding methods.  We invite researchers to build on these efforts\footnote{Our code can be found here: \url{https://github.com/ngillani/dynamic-word-emb}.} to further shed light on which language models are most appropriate for capturing changing perceptions and attitudes in our digital public sphere.

\section*{Acknowledgments}

We thank Doug Beeferman, Prashanth Vijayaraghavan and David McClure for their valuable input on this project.  We also thank Nikhil Garg for sharing prior results from his work. 

\bibliography{naaclhlt2019}
\bibliographystyle{acl_natbib}



\section*{Supplementary material}

# Corrigendum for "Simple dynamic word embeddings for mapping perceptions in the public sphere"

## Abstract

Upon revisiting code for this paper to use for another project, we identified several errors that, once corrected, invalidated the results presented in the original paper. This corrigendum describes the errors and seeks to explain why our original results appeared sensible.

## 1 Identified errors and shortcomings

The discovered errors and shortcomings can be classified into the following categories: i) computing linguistic bias, ii) data representation, and iii) model and training.

### 1.1 Linguistic bias analysis

1. We initially revisited some of the code I wrote for this project to use it for another project. Upon doing so, we noticed a simple programming error: when averaging over a sent of reference word list vectors, we failed to specify which axis to compute the mean over. This operation, then, returned a scalar value (instead of an average vector value) that we then subsequently used in the calculations. Because the scalar was a NumPy scalar, the calculation proceeded without error.

2. The dynamic embedding for word $w$ is defined as the sum of its attribute value-invariant vector and its attribute value-conditioned offset vector. To compute this sum when producing the linguistic bias results, we used the "+" operator—however, because these vectors were represented as Python lists (not NumPy arrays), this resulted in a vector concatenation instead of addition.

3. We did not normalize word vectors before inputting them into the relative norm distance bias calculation described in the main paper. Had we used NumPy's cosine distance implementation, these vectors would have been normalized as a part of the distance computation. Indeed, we found that using non-normalized vectors yielded a bias metric that was largely uncorrelated with an analogous bias metric computed using cosine distances; normalizing the vectors, however, led to high (almost perfect) correlation between these two measures, as also reported by (Garg et al., 2018). Unfortunately, we did not perform this sanity check when writing the original paper.

### 1.2 Data representation

1. The GitHub repo[1] we forked (a PyTorch implementation of Paragraph Vectors (Le and Mikolov, 2014)) to write the code for this project was subtracting one from the index returned by the PyTorch datastructure used to build the vocabulary (namely, the "stoi"—i.e., "string-to-index" dictionary mapping produced when creating a vocabulary from a TabularDataset[2] in PyTorch). This was, presumably, to avoid learning an embedding for the 0th element of the "stoi" mapping, or the "`<unk>`" (unknown) token. We did not inspect the repo's code closely enough to recognize this, and so, when looking up vectors in the model's learned parameters to compute linguistic bias, we failed to subtract one—thereby looking up the wrong vectors for words. We've updated the model to include the 0th element of the "stoi" in our model, which simplifies indexing so that the indices in the word embedding paramater

---

[1] https://github.com/inejc/paragraph-vectors.
[2] https://torchtext.readthedocs.io/en/latest/vocab.html.

matrices match the indices in the "stoi" mapping for the same words.

2. Because the aforementioned "stoi" mapping is implemented as a Python default dictionary, it does not throw an out-of-vocabulary (OOV) error when seeking the index for an OOV word—instead, it returns the embedding for the 0-index token (again, "¡unk¿"). Hence, in some cases when computing linguistic bias using words that weren't in the training set, the "stoi" mapping returned an incorrect vector. We've modified our code to skip OOV altogether instead of computing with them using an incorrect vector.

3. The repository we forked also implemented its own data generator to produce and serve up batches for training. These batches, however, did not contain a randomly shuffled set of word contexts from the training corpus; instead, they were generated by sequentially stepping through the corpus. This may have increased the likelihood (especially for long texts in our training data) that the training procedure learned unwanted correlations between words and the documents they were contained in, instead of focusing primarily on learning the semantic contexts in which they occurred. We updated this method to generate batches containing random word contexts drawn from the training data, leaning upon PyTorch's DataLoader class[3].

### 1.3 Model and training

1. The base repository also initialized the output layer parameters to a matrix of zeros. This may not have been the source of any bugs, but for the sake of consistency, we changed this to be a matrix of random numbers drawn from a standard normal distribution (which is how the repository implemented the other parameters in the model).

2. Previously, we averaged over the parameters of context words before multiplying them through the output layer parameters. Again, this might not have been the source of errors, but we changed this to take the sum over the parameters instead (just in case the average washed over important word-level semantic differences).

3. Initially, we applied l2 regularization to *all* model parameters, instead of only the matrix of attribute value-specific word offsets. Intuitively, we want our attribute-invariant word vectors to learn as much as possible about a given word's semantics across the entire corpus, deferring to the attribute-value specific offsets for that word only "as much as necessary" to properly represent the attribute value-specific meaning of that word. To achieve this, we updated the training procedure to apply l2 regularization only to the attribute value-specific word offsets—which seeks to satisfy the "as much as necessary" desideratum mentioned above. We also reduced the value of the regularization hyperparameter $\lambda$, which we initially set to 1e-2 (a relatively large value that intuitively would make the offset norms very small and perhaps lead to model underfitting), to 1e-5—more in line with recommendations from other deep learning research (Brownlee, 2018). Qualitatively, this led to large improvements in commonsense word similarity tests for the inferred word vectors (which we will describe more in the following sections).

## 2 Changes to initial results

The results presented in the updated version of the paper were computed after making the above corrections. Our updated results differ from those reported in the original paper in the following ways:

1. In the original paper, our model's gender and ethnic occupation bias scores closely matched those from (Garg et al., 2018) and ground truth workforce participation rates. In the updated paper, our model's scores seem to align reasonably well with (Garg et al., 2018) and ground truth for gender occupation bias, but not ethnic occupation bias.

2. The original paper also revealed a stark difference in the bias scores produced by our dynamic model and its non-dynamic counterpart when applied to the time series radio data. In particular, our model appeared to produce "smoother" bias scores that seemed to highlight a prominent news story during the timeline, whereas the bias scores

---
[3]https://pytorch.org/docs/stable/data.html.

produced by the non-dynamic model were much noisier. Our interpretation was that the smooth score trends were perhaps a result of the regularization and data-pooling enabled by the dynamic model. Our update results show that both models' resultant scores are comparably "non-smooth", and both appear to track the news cycle in some way. As discussed in the updated paper, however, the raw values for bias scores produced by our dynamic model still appear to more accurately depict what are likely to be biases against refugees in predominantly conservative talk radio shows.

3. Finally, in the original paper we also applied our dynamic model to learn geography-conditioned attributes. This was the most speculative analysis of the paper, and the main result was a marginally insignificant negative correlation between a city's talk radio bias against refugees and the number of refugees admitted by its containing state in 2017. After updating our model and results, there is still no significant correlation. Since this was a speculative analysis to begin with that rested upon a stronger validation of the model's results that existed in the initial version of the paper, we have left it out of the updated version.

## 3 Interpreting originally-reported results

In light of the errors highlighted above, one glaring question is: why do the results reported in the original paper seem sensible? We begin by challenging the assertion that the original results were ever actually sensible to begin with. A critical oversight in our initial analysis was a failure to perform simple sanity checks to understand our model's behavior. In particular, we did not conduct several common sense word association tests to check the distance between inferred vectors for words that we would expect to be semantically similar / dissimilar.

We conducted a series of distance calculations between words we would expect to have similar vector representations (averaged across attribute values). Figure 1(a) illustrates these results after using our original model / code and training on COHA. For comparison purposes, figure 1(b) shows the results produced after correcting the errors highlighted above. From this, we can see that

Figure 1: Commonsense word association tests for words selected across each of our corpora (where semantic distances are averaged across attribute values—i.e. decades, days, and cities for each of the models highlighted above, respectively). The results from the original model do not appear to make sense, while the results from the corrected model do.

the original model do not yield the behavior one might expect, whereas the corrected model does.

Another issue with the original results—and in particular, the comparisons we made to the methods and data presented in (Garg et al., 2018) (while also using COHA)—is that the raw magnitudes of the decade-by-decade bias scores we reported actually had the opposite signs one might expect. In particular, for most decades, the results suggest that words describing Women and Asians were actually *more* likely to be associated with occupation-related words—whereas (Garg et al., 2018) (and historical labor participation rates for these groups) actually suggest the opposite. However, we projected my bias scores, those from (Garg et al., 2018), and the actual historical labor participation rates into the same coordinate space by subtracting the mean and dividing by the standard deviation for each so as to be able to visualize them on the same graph. This data standardization procedure produced results that, directionally, appeared to check out. Unfortunately, we didn't double check to make sure the raw underlying values actually matched intuition or historical trends.

It is still unclear, however, why the directional trends depicted in our sanity checks and the results reported in (Garg et al., 2018)—both after standardization—appeared to align so closely. Perhaps one reason is that the model learned some other feature of the data that correlated with the bias scores / labor participation rates presented in that paper. In particular, there has been some prior work illustrating how the norms of word2vec vec-

tors can encode information about the frequency of that word: words that occur frequently and in a consistent semantic context tend to have larger norms than frequent, common words (like "it" or "the", which likely occur in varied semantic contexts) or infrequent words (Schakel and Wilson, 2015). Since we did not normalize our vectors before using them to compute the relative norm distance-based bias scores, it is possible that the bias scores themselves—which are a function of the vector norms for several pre-selected words—might actually be tracking the frequencies of those words more than anything.

Indeed, there are several pieces of evidence that suggest this might be the case. The first few pieces are actually in the original paper itself—starting with where we wrote "Computing the correlation between daily bias scores and the number of mentions of the keyword refugee across stations yields r = 0.56, p < 0.001, suggesting that additional discourse about refugees tends to be biased against them." Upon taking another look, it is actually the term "refugees", not "refugee", that has this correlation. Nevertheless, while "refugees" is only one of the seven terms[4] we use to define the group to compute linguistic bias with respect to—and the only one that has a correlation with the bias scores of this magnitude—it is not out of the realm of possibility that the bias scores might be tracking the norm of this vector. Similarly, when discussing the geographic bias results, we mentioned "Interestingly, there is a weak negative, though marginally insignificant, correlation between the level of bias per city and the number of refugees the city admitted in 2017." It turns out that in this case, there is also actually a weak, though statistically insignificant, negative relationship between each city's bias score and the number of times "refugees" is mentioned across the radio shows per city (r = -0.1, p = 0.43). However, there is also a weak *positive*, marginally significant correlation between the number of refugees each city admitted in 2017 and the number of times "refugees" is mentioned on its radio shows (r=0.22, p < 0.05)— meaning the originally-reported correlation between bias levels per city and refugee admission numbers may actually be reflecting how frequently "refugees" are mentioned on radio.

When we look to our COHA-based comparisons with the results in (Garg et al., 2018), we see some similar trends. For example, the correlation between the per-decade total frequency of the 75 occupation-related words used to compute bias with respect to, and the per-decade gender-occupation bias scores, is 0.63. For the Asian-White occupation bias analysis, the correlation with the per-decade frequency of all occupation-related words is much lower—0.21—but the frequency of several individual words is still highly correlated with the decade-by-decade bias scores[5]. Of course, these correlations are computed over a very small number of data points (in particular, nine—one per decade), so should be taken with a grain of salt. Indeed, this "small data" may also help explain why the original validations against (Garg et al., 2018) seemed to check out: with such few data points, finding spurious correlations entirely was never out of the realm of possibility.

Of course, one question remains: if one of the errors in the original code was that the word vector look-ups into the trained model were off by one due to the indexing error described in section 1, then why should there be any relationship at all between the frequency of words in our word lists, their norms, and the computed bias scores? Surely the norms I computed as a part of the bias scores aren't even for vectors that correspond to the intended words. One explanation could be that the "stoi" (string-to-index) dictionary described earlier in section 1 that maps words to indices (which, in turn, indicate the locations in the model to look up the word's corresponding vector) is actually built using a list of words in the vocabulary that have been sorted by frequency. This means that even if indexing is off by one, it still retrieves a vector for a word that occurs with similar frequency as the intended word—perhaps explaining how its vector norm, and therefore by extension, resulting bias scores could be similar.

# References

J. Brownlee. 2018. How to use weight decay to reduce overfitting of neural network in keras. https://bit.ly/3mwU8Oj. Accessed: September 19, 2020.

---

[4]The others are "refugee", "asylum", "migrant", "migrants", "immigrant", "immigrants"

[5]e.g. the individual per-decade frequencies of "judge", "retired", "designer", "cook", "attendant", "teacher", "nurse", "artist", "athlete", "psychologist", "architect" all have a correlation with the per-decade Asian-White occupation bias score of 0.7 or higher



\end{document}